# BALL LIGHTNING AND PLASMA COHESION

John J. Gilman
Materials Science and Engineering
University of California at Los Angeles
Los Angeles, California 90095

## Abstract

The phenomenon of ball lightning has been observed for a long time, but the nature of these luminous balls has been unknown.  It is proposed here that they consist of highly excited Rydberg atoms with large polarizabilities that bind them together. Thus the cohesion of the balls comes from photon exchange forces (London dispersion forces) rather than the more usual electron exchange (chemical) forces.  The cohesion in plasmas generated at the back faces of detonating explosives has a similar basis. Estimates are given to justify this interpretation.

For centuries, luminous plasma spheres have been observed during thunder storms in addition to linear lightning discharges.  The literature regarding these "lightning balls" has been extensively reviewed in a recent special issue of the Transactions of the Royal Society of London (2002), and an extensive review of their characteristics was published about a decade ago (Turner, 1994).  However, no convincing model of the structure, and properties, of these remarkable spheres has emerged from this review.   In particular, their considerable cohesion is not explained, although numerous proposals have been made.

Cohesive plasmas have also been observed in association with the detonation of explosives.  Highly luminous "plasma-discs" have been photographed emerging from the back surfaces (the side opposite the initiator) of detonating explosives. First observed by Cook (1958), they were later observed by others: Davis and Campbell (1960); Fortov, Musyankov, Yakushev, and Dremin (1974); and Zernow (2001). These authors found that the plasmas persist long enough to travel in air for a few meters (of order microseconds). The lifetimes of lightning balls may be much longer; 10-100 seconds (Turner, 1994, p.96). Laboratory microwave plasmas persist for 200 ms. (Brandenburg and Kline, 1998).

Cook also observed that the plasmas have enough cohesion to turn sharp corners (180 deg.) when guided by glass tubing; indicating that the plasma has little shear stiffness. Later, Zernow (2001) observed that when the plasmas created by explosives struck aluminum witness plates they gave up enough energy to melt the surface. This effect verfies their approximate densities. But it probably has more to do with the kinetic energies of the total number of high speed particles than with the recombination energies of the ions. Cook (1958) and his colleagues proposed that the explosives generated plasmas are analogous to dense metallic "glasses", but Fortov, Musyankov, Yakushev, and Dremin (1974) proved that the plasmas induced by explosives are of relatively low density (about 10X the density of air); so they cannot be metallic.

Related plasma "balls" (clouds of excitons) in Ge crystals have also been excited by irradiating the crystals with intense coherent light (Kittel, 1996). In this case the binding energies of the excitons are about 2 meV, and the concentation is about 2 X $10^{17}$ / cc.

An important clue to the cohesion of these various plasmas is their luminosity. This indicates that the particles within them are in excited Rydberg states. Ordinary excitations would not be expected to yield substantial cohesion, but the van der Waals (and Casimir) forces between the atoms and molecules increase with the squares of their polarizabilities. The latter, in turn, increase with squares of the principal quantum numbers of the excited states. This is because the polarizability is proportional to the volume of an excited species (whose radius is determined approximately by the square of the principal quantum number). Rydberg atoms with principal quantum numbers as high as n ≈ 1100 have been observed in laboratory experiments (Dunning, 1995). Large polarizabilities have also been measured (van Wijngaarden, 1999).

The mass densities of the plasmas in question are relatively small as indicated by their translucence, and by the direct measurements of Fortov et al. (1974). Their ionization fractions are about 10%. Therefore, the point particle-particle Coulombic cohesion in them is negligible. It is the London dipole-dipole interactions between the Rydberg atoms that yield significant cohesion. This consists of two principal terms: one is a repulsive overlap term; and the other an attractive dipole-dipole term (both are induced dipoles). These terms differ in range, and when averaged over a plasma sphere yield net cohesion. This may be thought of as Lennard-Jones cohesion since all of the interaction forces are positive, and of long range.

The observed diameters of ball lightning spheres are summarized by Bychkov, Bychkov, and Abrahamson (2002). They range from 0.02 to 1.5 m. with an average of 0.25 m. Take 0.2 m. as typical, or a radius of 10 cm. This corresponds to the maximum possible Rydberg orbital, $r_R$ which is given approximately by the Bohr expression (Gallagher, 1994):

$$r_R = (n^2\hbar^2) / (Zq^2m) = 053 \times 10^{-8} \text{ cm.} \quad \text{(for } n = Z = 1\text{)} \quad (1)$$

so for air with Z (nitrogen) $\approx$ 7; q = electron charge; m = electron mass; $\hbar$ = Planck's constant / $2\pi$; and $r_R$ (max.) = 10 cm.; this expression yields $n \approx 10^4$. But the average value of $n^2$ is $n^2$ /3, so the average $r_R$ is about 3 cm. At this radius the binding energy of an electron to its positive ion is very small, but the polarizabilty, $\alpha \approx 4\pi r_R^3/3$ of the average Rydberg atom is very large. It is about $40\pi$ cm$^3$ which is about $10^{25}$ times larger than that of an individual nitrogen atom in its ground state. The corresponding interaction energy between two such Rydberg atoms (with $\alpha$ = polarizability, I = ionization energy = $-q^2 / 2r_R$ , and d = distance between them) is about:

$$u = - [(3/4)\alpha^2 I] / \epsilon_p d^6 \quad (2)$$

where q is an electron's charge (Israelachvili, 1991).

The interaction between any two Rydberg atoms immersed in a plasma ball is screened by the polarizabilities of the other atoms. Therefore, it is expected that the average interaction energy will be reduced by the effective dielectric constant, $\epsilon_p$ of the plasma (where $\alpha$ is the average polarizability):

$$\epsilon_p = 1 + 4\pi N\alpha \approx 14 \quad (3)$$

since $N\alpha \approx 1$

**Densities of Lightning Balls**

Since lightning balls are observed to travel horizontally for considerable distances, and they are translucent, their densities must be comparable with that of air;

namely, about $1.185 \times 10^{-3}$ g/cc. at 25°C.  A typical volume per ball is $4.19 \times 10^{3}$ cc. (10 cm. radius), so the weight is 4.96 g.  Assuming air to be 20% oxygen, and 80% nitrogen, there are about 0.17 mols / ball.; hence, $2.08 \times 10^{23}$ particles / ball.  This yields an average inter-particle spacing of about 27 Å.  The ion density is less than the particle density, of course (Anderson, 1981)

To get an upper limit for the cohesive energy, since the dipole-dipole interaction energy always has the same sign, the pair energies (given by Equation (2)) can simply be added up.  The degree of ionization in lightning balls is not known, but in the plasma disks generated by explosives Fortov et alia (1974) measured it to be about 10%.  Since the luminosity in the two cases is about the same, it will be assumed that the degree of ionization is the same.   This is only a guess with a weak basis, but it yields an ion density of about $1.12 \times 10^{-4}$ g / cc.  With the average molecular weight of air being 29 g / mol., the particle density is $2.3 \times 10^{18}$ ions / cc., so the average distance between ions is 75 Å.  This means that the average polarizability is about $5.3 \times 10^{-20}$ cc.  The ionization energy of nitrogen is 15.6 eV (one eV = $1.6 \times 10^{-12}$ erg.),

Substituting these numbers into Equation (2):

$$u = 0.014 \text{ eV}$$

which is roughly 1% of the cohesive energy per atom in a metal, and therefore reasonable.

The shear modulus, G, an indicator of resistance to shear deformation, can be estimated by comparing the deformation of a plasma ball caused by an external electric field (polarizability = $\alpha$) with that caused by mechanical strain.  The resulting relation is:

$$G = [3/4\pi][q^2 / \alpha d] \qquad (4)$$

where d is the average spacing between ions and electrons, q = electron's charge = 4.8 x $10^{-10}$ esu. (Gilman, 1997). This gives G = 1.4 atm. Which is consistent with the deformability of the lightning balls. They have been reported to pass through openings smaller than themselves (Turner. 1994)